\documentclass{article}

 \usepackage[creativeai, final]{neurips_2025}

\usepackage{float}
\usepackage[utf8]{inputenc} 
\usepackage[T1]{fontenc}    
\usepackage{hyperref}       
\usepackage{url}            
\usepackage{booktabs}       
\usepackage{amsfonts}       
\usepackage{nicefrac}       
\usepackage{microtype}      
\usepackage{xcolor}         
\usepackage{graphicx}

\title{\textit{Sound Clouds}: Exploring ambient intelligence in public spaces to elicit deep human experience of awe, wonder, and beauty}

\author{
  Chengzhi Zhang\textsuperscript{1} 
  \hspace{1mm}  Dashiel Carrera\textsuperscript{1,2,†} 
  \hspace{1mm}  Daksh Kapoor\textsuperscript{1,†} 
  \hspace{1mm}  Jasmine Kaur\textsuperscript{1,†} 
  \hspace{1mm}  Jisu Kim \textsuperscript{1,†} \\
   Brian Magerko\textsuperscript{1} \hspace{1mm}  \\
  \textsuperscript{1}Georgia Institute of Technology, Atlanta, GA 30309, USA \\
  \textsuperscript{2}University of Toronto, Toronto, ON, Canada \\
  \texttt{\textsuperscript{1}\{czhang694, dkapoor30, jkaur47, jisu.kim, magerko\}@gatech.edu}, \\
  \texttt{\textsuperscript{2}dcarrera@dgp.toronto.edu}\\
  \vspace{2mm}
  \small{\textsuperscript{†}Equal contribution}
}

\begin{document}

\maketitle
\vspace{-20pt} 

\begin{figure}[ht]
    \centering
    \includegraphics[width=0.8\textwidth]{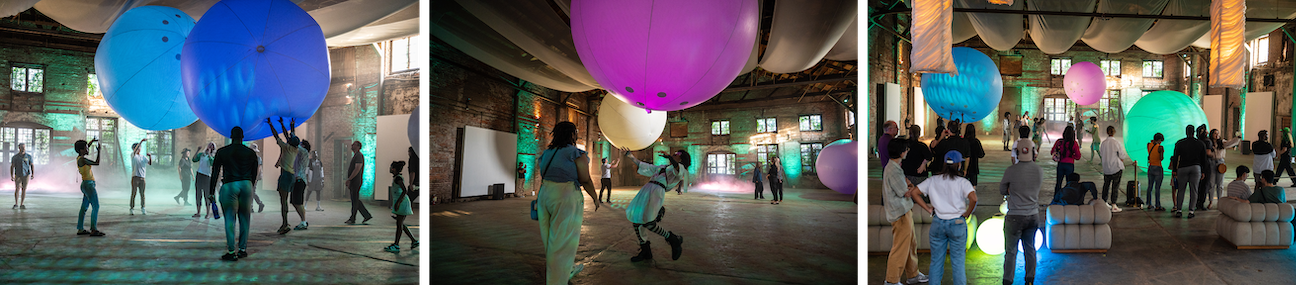}
    \caption{Participants Interacting with \textit{Sound Clouds} on the Live Exhibition Day}
    \label{figure:people_interacting}
\end{figure}

\begin{abstract}
  While the ambient intelligence (AmI) systems we encounter in our daily lives, including security monitoring and energy-saving systems, typically serve pragmatic purposes, we wonder how we can design and implement ambient artificial intelligence experiences in public spaces that elicit deep human feelings of awe, wonder, and beauty. As a manifestation, we introduce--\textit{Sound Clouds}, an immersive art installation that generates live music based on participants' interaction with several human-height spheres. Our installation serves as a provocation into future ambient intelligence that provokes, not limits, the future possibilities of AmI.   
\end{abstract}

\section{Introduction}

Ambient intelligence (AmI), as a subset of artificial intelligence, has started to play an increasingly inseparable role in our daily technological encounters. While there are varied definitions of AmI, one widely adopted definition from ~\cite{aarts2006true} for ambient intelligence is ``developing technology that will increasingly make our everyday environment sensitive and responsive to our presence''. AmI generally possesses the characteristics of being Sensitive (S), Responsive (R), Adaptive (A), Transparent (T), Ubiquitous (U), and Intelligent (I)~\citep{cook2009ambient}. While most of the ambient intelligence serves a pragmatic purpose, such as a) the \textit{Nest thermostat}~\citep{hernandez2014smart} for achieving domestic energy-saving, \textit{b) ambient influence} for promoting positive behavior change like taking the stairs~\citep{rogers2010ambient}, and c) \textit{LAICA} for enhancing public security~\citep{cucchiara2005ambient}, etc., there has been little exploration in other kinds of human experience with ambient intelligence. Therefore, we ask the following overarching research question:
\textit{How can we design interactions with ambient artificial intelligence in public spaces to elicit deep human feelings of awe, wonder, and beauty?}

We introduce \textit{Sound Clouds}, an exploratory ambient intelligence art installation that consists of several larger-than-human-height (8-12' in diameter) floating spheres that were filled with helium and air to (approximately) reach neutral buoyancy. Participants can use any or all of their bodies to interact with the spheres while they control real-time music that is generated from the movement and location of the spheres. There is no instruction or explanation of the system given to participants--curiosity alone guides their interactions and exploration.

\section{\textit{Sound Clouds}}

\begin{figure}[ht]
    \centering
    \includegraphics[width=0.4\textwidth]{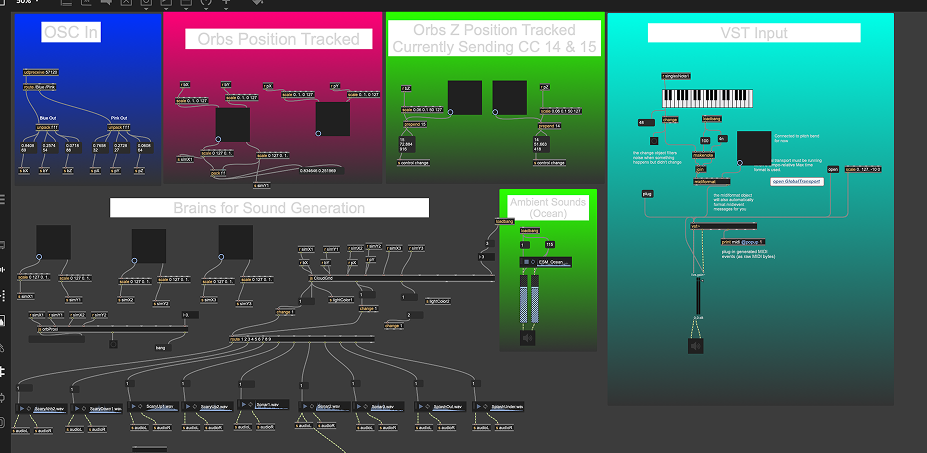}
    \caption{Screenshot of the \textit{Sound Clouds} Music Generation Logic Implemented in Max/MSP}
    \label{figure:maxScreenshot}
\end{figure}

\textit{Sound Clouds} is an interactive ambient intelligence environment comprising several human-height Polyvinyl Chloride (PVC) spheres (with diameters from 1.2 m to 3 m). It was exhibited inside the building of a renovated factory with interior dimensions of 18 m (width) $\times$ 42 m (length) $\times$ 9 m (height). \textit{Sound Clouds} was experienced by more than 200 participants during our two-hour public exhibition. Inside the building, we used dry ice and fabric canopies hanging on top of the ceiling to create a serene feeling and projected ``waves'' of light on the ground to create an underwater theme. We mounted a GoPro\footnote{GoPro~\url{https://gopro.com/en/us}} camera on the ceiling facing top-down to capture the live video input of the space. The video feed is routed via Wi-Fi to a central-processing Mac computer, which runs a computer vision (CV) model for object detection to obtain spheres’ location and diameter. The diameter is used to determine the sphere's height.
The central-processing Mac computer used the ESP-NOW protocol~\footnote{ESP-NOW~\url{https://github.com/espressif/esp-now}} to send the light signal to each ESP32 board embedded inside the sphere. According to the spheres' locations, \textit{Sound Clouds} generates live music in Max/MSP (see Figure~\ref{figure:maxScreenshot}) software and changes the light of the spheres accordingly. We allowed at most 20 participants to interact simultaneously. 

\begin{figure}[ht]
    \centering
    \includegraphics[width=1\textwidth]{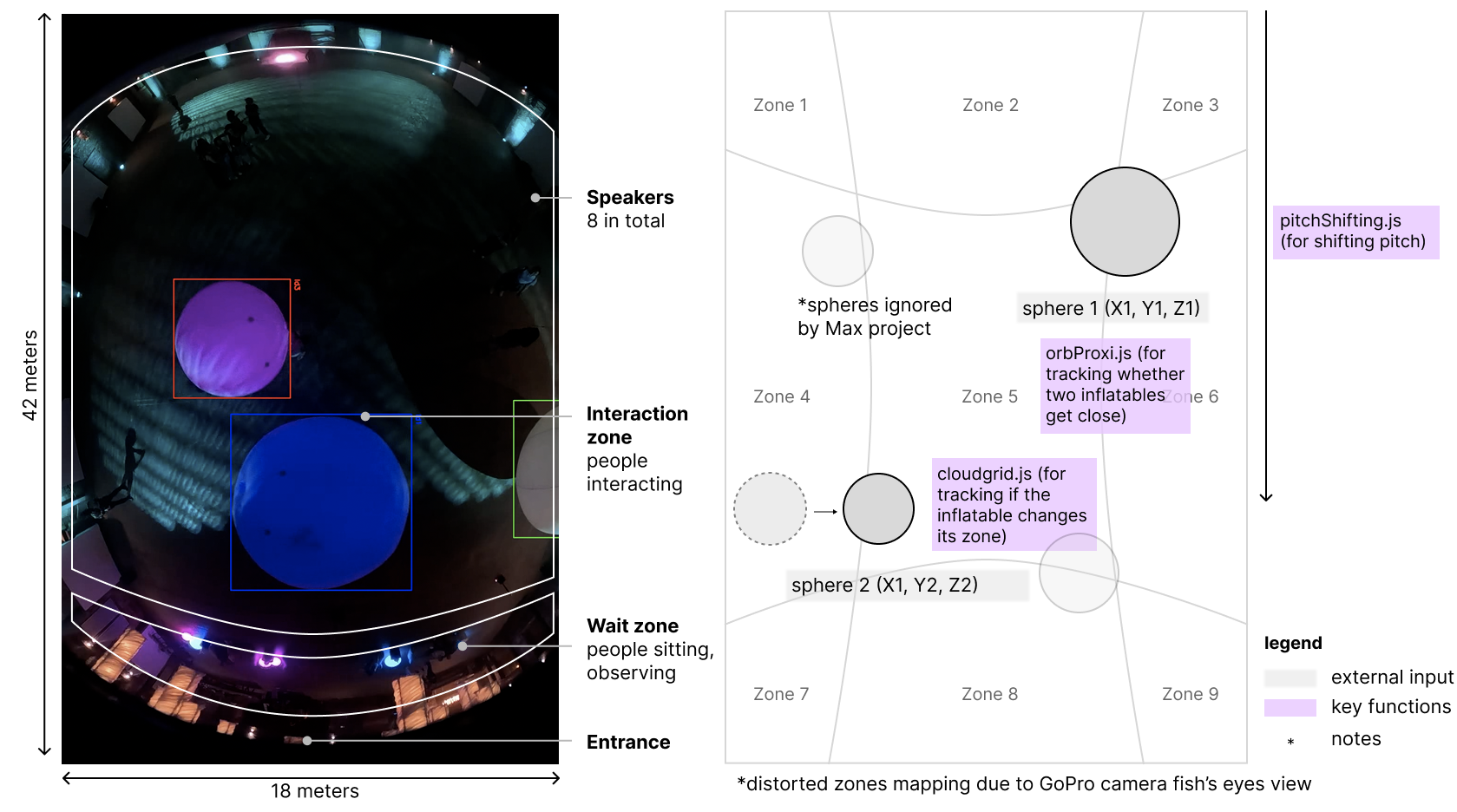}
    \caption{\textbf{Left:} Top-down View of the Exhibit Space (View from the Go Pro Camera), \textbf{Right:} Top-down Schematic of the Max Music Generation Project}
    \label{figure:topDown}
\end{figure}

\subsection{The roles of AI and ML}

We used computer vision (CV) for tracking spheres using a fine-tuned YOLO12n model~\footnote{YOLO12n \url{https://docs.ultralytics.com/models/yolo12/}}. The music generation is rule-based through Max/MSP~\footnote{Max/MSP Wikipedia~\url{https://en.wikipedia.org/wiki/Max_(software)}} software. Although the CV model tracked all the spheres, we used the parameters of two spheres for music generation. We looped an ocean-like sound as the background, depicting the ethereal, mysterious atmosphere. We divided the video feed space into nine zones and used three key scripts for sound generation (see the description below and Figure~\ref{figure:topDown}, Rigt). Due to the fish-eye's view of the GoPro wide-angle lens, the vision is lightly distorted (see Figure~\ref{figure:topDown}, Left). The key scripts for interaction are:

\begin{itemize}
    \setlength{\itemsep}{0pt}
    \setlength{\parskip}{0pt}
    \setlength{\parsep}{0pt}
    \item \textbf{cloudGrid}: One sphere enters a new zone, triggering a corresponding and distinguished pre-recorded sound. 
    \item \textbf{orbProxi}: Two spheres close enough to each other below a threshold trigger a sound.
    \item \textbf{pitchShift}: The depth of one sphere shifts the pitch of the background sound.
\end{itemize}

\section{\textit{Sound Clouds} and Humanity}

While the ambient intelligence usually serves a pragmatic purpose, \textit{Sound Clouds} elicits deep human experience of awe, wonder, and beauty. Awe, as a specific self-transcendent emotion, is characterized by ``the perception of being in the presence of something vast that the individual does not immediately understand''~\citep{stellar2017self}. As a rarely-experienced emotion, experiencing awe has been found to prompt feelings of interconnectedness~\citep{shiota2007nature}, motivate commitment to social collectives~\citep{keltner2003approaching}, and improve perception of well-being~\citep{rudd2012awe}. Thus, experiencing awe can contribute to the exploration of humanity aspects of AI. Participants shared their feelings of awe during the post-interaction interview, including: 

\vspace{-8pt} 
\begin{quote}
    ``...the ambiance and the whole vibe of the place was very \textbf{peaceful and calming}''\newline
    ``just like, jaw on the floor...it felt almost like \textbf{dream-like}''\newline
    ``I think that that's one of the things that we lose in childhood, from childhood, as adults, we don't play because the proportions are off. That \textbf{sense of wonder and awe} are different.''\newline
\end{quote}
\vspace{-15pt} 

Following the research thread of \textit{The Aware Home}~\citep{abowd2002aware}, an AmI research space supporting the elderly's dignified aging, and \textit{Smart Kitchen}~\citep{bell2002designing}--which prioritizes ``experience, affect, and desire'' but not just efficiency, our exhibit serves as a provocation for future AmI systems that enrich--not replace--humanity.
 
\section{Author Bibliography}
\begin{itemize}
    \setlength{\itemsep}{0pt}
    \setlength{\parskip}{0pt}
    \setlength{\parsep}{0pt}
    \item Chengzhi Zhang is a PhD student at Georgia Tech with a research focus on AI literacy~\citep{zhang2025generative} and the intersection of AI and UX~\citep{zhang2023generative}. 
    \item Dashiel Carrera is a Visiting PhD Researcher at Columbia University, a novelist, and a sound media artist whose research and art practice explore AI's impact on the Arts~\citep{10.1145/3698061.3726916}.
    \item  Daksh Kapoor is a graduate student at Georgia Tech with a research focus on embodied interaction and experience design, with an emphasis on how users interpret, engage with, and move within immersive and interactive systems. 
    \item Jasmine Kaur is a graduate student at Georgia Tech with a research focus on human-computer interaction and speculative design with a special emphasis on creative AI technologies~\citep{kaur_bringing_2025}.
    \item Jisu Kim is an M.S. student and Graduate Research Assistant at Georgia Tech with a research focus on responsible AI~\citep{lee2024one} and human-AI collaboration, specifically how AI can augment--rather than replace--human values.
    \item Brian Magerko is a Regent's Professor of Digital Media and Director of Graduate Studies at Georgia Tech and has contributed broadly over 3 decades to creativity, cognition, and computing research with a focus on human/AI co-creativity.
\end{itemize}

\bibliographystyle{abbrvnat}

\end{document}